\documentclass[useAMS,usenatbib]{mn2e}

\usepackage{amssymb}
\usepackage{amsfonts}
\usepackage{mncite}
\usepackage{epsfig}
\usepackage{psfig}

\newcommand{\sub}[1]{_{\rmn {#1}}}
\newcommand{\HII}{H\,\textsc{ii}}

\defcitealias{kenyon91}{KH91}

\begin{document}

\title[Star formation bursts in isolated spirals]
{Star formation bursts in isolated spiral galaxies}
\author[C. Clarke and D. M. Gittins]
{C. Clarke and D. Gittins \\
Institute of Astronomy, Madingley Road, Cambridge, CB3 0HA}

\maketitle

\begin{abstract}
We study the response of the gaseous component of a galactic disc
to the time dependent potential generated by
N-body simulations of a spiral galaxy.
The results show significant variation of the spiral structure of
the gas which might be expected to result in significant fluctuations
in the Star Formation Rate (SFR). 
Pronounced {\it local} variations of the SFR  are anticipated in all
cases. Bursty histories for the {\it global} SFR, however,
require that the mean surface density is much less (around an order
of magnitude less) than the putative threshold for star formation.
We thus suggest that bursty star
formation histories, normally attributed to mergers and/or tidal interactions, 
may be a normal pattern for gas poor {\it isolated} spiral galaxies.

\end{abstract}


\section{Introduction}
  The local Universe furnishes many illustrations of the fact that
vigorous star formation is highly localised in both space and time. The
most spectacular examples, circumnuclear starbursts can, over timescales
of $\sim 10$ Myr, sustain star formation rates comparable to entire
galaxies from a region less than a kiloparsec across (Lehnert
 and Heckman 1996).  Equally intense,
but more spatially distributed, is the current burst of star formation
(and Super Star Cluster formation) in the Antennae Galaxy (Whitmore
 et al 1999). Star
formation in the Antennaa is self-evidently triggered by 
 a major
 merger;  there is also ample evidence, however, 
 that bursts of star and cluster formation can be triggered by
 interactions with small satellite galaxies
 (i.e. by minor mergers; see, for example, Homeier and
 Gallagher 2002).

  In isolated galaxies, however, it is usually assumed that the global star
formation rate (henceforth SFR) shows little time variation although,
in the case of disc galaxies, star formation is clearly spatially
localised along spiral arms. This assumption (of a roughly constant
quiescent SFR) is partly based on the expectations of spiral density wave
theory in which the disc is susceptible to non-axisymmetric
instabilities that generate a {\it long lived} pattern of spiral arms
(Bertin et al 1989).
Thus whereas the SFR may vary locally as material is compressed
by the passage of spiral density waves, the pattern is invariant in a
frame corotating with the spiral disturbance and therefore the global
SFR is expected to be roughly constant.
  N-body simulations of self-gravitating discs however paint a very different
picture of the development and longevity of spiral structures. For
example, Sellwood and Carlberg 1984 performed N-body simulations of a
stellar disc subject to both self-gravity and to a rigid halo potential.
They found that the spiral structures that developed in their simulations
were {\it transient} features which formed and re-formed on a timescale 
comparable 
with the galactic rotation period. (See also Huber and Pfenniger 2001, Gerritsen and Icke 1997 and Bottema 2003 
for further simulations producing transient, regenerative
spiral structures).
The development and dissolution of such
structures means that at a given location the gas is no longer subject to
periodic variations in the potential, as in classical spiral density
wave theory, and thus one would expect irregular
variations in the resulting SFR.
Likewise, the pattern of evanescent structures may not even give rise
to SFRs that are {\it globally} constant, an impression that is
strengthened by inspecting snapshots of the N-body simulations
(see Figure 5 of Sellwood and Carlberg 1984), which suggest that potential
troughs are much more pronounced at some epochs than others.

  In this paper we undertake a preliminary investigation of the
response of a galaxy's gaseous component to such irregular potential
variations. In this simple approach, we neglect the self-gravity of
the gas, and its potential role in amplifying the structures in
the N-body simulations, and instead study the response of  non-self gravitating
isothermal gas. In order to make qualitative statements about the spatial
and temporal variations of the resulting SFR we have to make some assumption
about how SFR depends on the surface density distribution. Here we simply
associate the SFR with the fraction of the gas mass that is instantaneously
compressed to a column density greater than some adjustable threshold
value, $\Sigma\sub{crit}$ (see Kennicutt 1989 and Martin and
Kennicutt 2001 for a discussion of such a star 
formation threshold in disc galaxies). Note that we do not
actually deplete the gas surface density in response to this nominal
star formation rate but that since the simulations extend only over a few
galactic rotation periods, this is not a major shortcoming.

 The structure of the paper is as follows. In Section 2 we describe the
 N-body and gas dynamical simulations on which the study is based, in
 Section 3 we describe the resulting gas structures and in Section 4
 interpret these time dependent structures in terms of a nominal
 star formation rate. Section 5 summarises our conclusions.

\section{Numerical Method}

\label{sec:numerical}
The calculations are based on the gravitational potential formed in an N-body
simulation of a spiral galaxy, which was performed by J.~Sellwood, who kindly supplied
the resulting data.  This simulation is described below in section 2.1. 
The two-dimensional potential
in the galactic plane was then applied as an external potential to an isothermal
hydrodynamical gas disc, using the two-dimensional PPMLR grid code  CMHOG, which
was provided by P.~Teuben.  This second calculation is described in section 
2.2.

\subsection{The N-body galaxy simulation}
\label{sec:nbody}
  The simulation from which the potential data are taken is a high resolution,
three dimensional version of those presented in Sellwood and Carlberg 1984.
J. Sellwood kindly supplied us with data on the potential variations
by re-running in three dimensions (and at greater resolution) the
timespan between $t=0$ and $t=4$ in Figure 2 of Sellwood and Carlberg.
In this simulation the stellar velocity dispersion is allowed to rise
due to heating of the disc by gravitational instabilities. After
several rotation periods, the rising velocity dispersion stabilises
the disc against non-axisymmetric instability and the Toomre
Q parameter (Toomre 1964) attains a limiting value of around $2$. At this
point (latter panels in Sellwood and Carlberg's Figure 2)
the spiral arms become broader and fainter.
However, Sellwood and Carlberg showed that a recurrent pattern of
transient spiral arms will persist indefinitely if the disc is
subject to a cooling agent that prevents the secular rise of
the stellar velocity dispersion. They argued that in a real galaxy,
accretion and dissipation in the gas phase would provide suitable cooling.
In their second set of simulations, however,
they instead mimicked cooling by the
continual injection of disc particles on `cool' (i.e. circular) orbits.
The resulting spiral patterns (Figure 5) are very similar to those seen
in the uncooled simulations over time $t=0-4$, but now persist over
many galactic rotation periods. In order to avoid the computational
complications involved by introducing a pseudo cooling prescription into
the calculations, we here just  use the uncooled simulation over
the timescale $t=0-4$, noting its similarity to the long term state
of the cooled calculations.

 The simulation described here consists of 1M particles,
whose gravitational field
is  evaluated over a cylindrical polar grid consisting of $225$
planes of $90 \times 128$ points.
The dimensionless galaxy model is scaled to real units by choosing a maximum
radius of 30 kpc and a rotation velocity at the peak of the rotation curve
of 220 km s$^{-1}$.
With this choice of units, the spacing between
 grid planes is $19$ parsecs, and the gravitational softening
 parameter (see Sellwood and Carlberg 1984) is $38$ parsecs.

The N-body calculation places particles in a surface-density distribution
corresponding to the chosen rotation curve,
while
 an additional rigid halo potential, corresponding
 to the same rotation curve, is added, which suppresses the bar instability
 (Ostriker and Peebles 1973). The disc and halo respectively comprise $30 \%$ and
 $70 \%$ of the total mass and the resulting rotation curve
 (in real units) is shown in Figure 1.
 The disc particles are spread vertically according to the
 isothermal sheet solution  of Spitzer (1942)  with scale parameter  (see
 problem $4-25$ of Binney and Tremaine 1987 for definition) $z0=77$ parsecs
 and a vertical velocity dispersion appropriate to such a system in
 equilibrium.
 The other components of the velocity dispersion are chosen so as to
 satisfy the Toomre stability criterion, i.e.\ to ensure
that $Q\sub{stars}=1$.  Spiral structure naturally emerges as the
simulation proceeds.

\begin{figure}
\centerline{\epsfig{figure=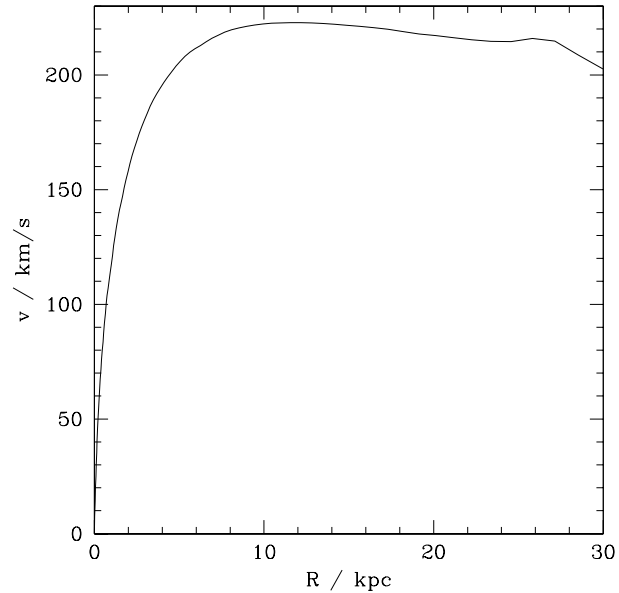,width=84mm}}
\caption{ Rotation curve of the N-body simulation from which
the gravitational potential was extracted}
\label{fig:sellwoodvel}
\end{figure}

\subsection{Two-dimensional Piecewise Parabolic Method}
\label{sec:cmhog}

The 2D grid code  CMHOG,  provided by P.~Teuben, uses the
Piecewise Parabolic Method
in the Lagrangian Remap formalism (PPMLR), and includes isothermal hydrodynamics
in polar coordinates.  The code is based on that described in Piner et al
1995;
for
a description of the Piecewise Parabolic Method, see Colella and
Woodward 1984. In the simulations conducted here, the grid consists
 of 512 angular by 328 (logarithmically spaced) radial zones, and
 constant pressure boundary conditions are applied at the disc inner
 and outer edge. 

An initially uniform gas disc is
subject to the potential generated by the N-body calculation, this
comprising a rigid halo component and a time dependent potential
produced by the particles in the N-body code.
The only remaining parameter to be adjusted is the sound speed\footnote{Here, the sound speed used is an `effective' sound
speed equal to the velocity dispersion of the ISM, a good approximation for the time
and length scales involved (Cowie 1980)}
in the gas, $a$; the calculation
was repeated with values $a=5$ km s$^{-1}$ and $a=8$ km s$^{-1}$.

 This calculation therefore represents the response of the gaseous disc in a 
galaxy
in which the mass is dominated by the stars.  The spiral structure, which arises
naturally in the stellar orbits, generates a corresponding response
in the gas. We note that in real galaxies, the gas surface density
declines roughly exponentially with radius, so that our use of
uniform initial surface density in the gas will weight the global response of the
gas towards conditions at the outside. In fact, we will find that the fluctuating
spiral structure is stronger in the inner disc, so that the fractional
variations in `star formation' that we present are lower limits
to what would occur in a galaxy whose gas disc was realistically
centrally concentrated.
\section{Numerical results}
\label{sec:results}
Surface density maps of the response of the initially uniform gas disc are shown
 in Figure 2
%\ref{sellwood5_rhopics} 
(for the case $a=5$ km s$^{-1}$).
The gas develops a complex structure of spiral shocks, showing multiple arms
and significant deviation from straightforward spiral structure.  Bifurcation
of shocks, bending and merging of spiral arms and substructure perpendicular
to the main shocks are all seen.  At larger radii, shocks become weaker and
are more tightly wound than those closer to the centre.  Very little qualitative
difference is visible between the calculations at different sound speeds; at 
lower
sound speed, the shocks in the outer part of the disc are somewhat sharper,
as expected.

\begin{figure}
\centerline{\epsfig{figure=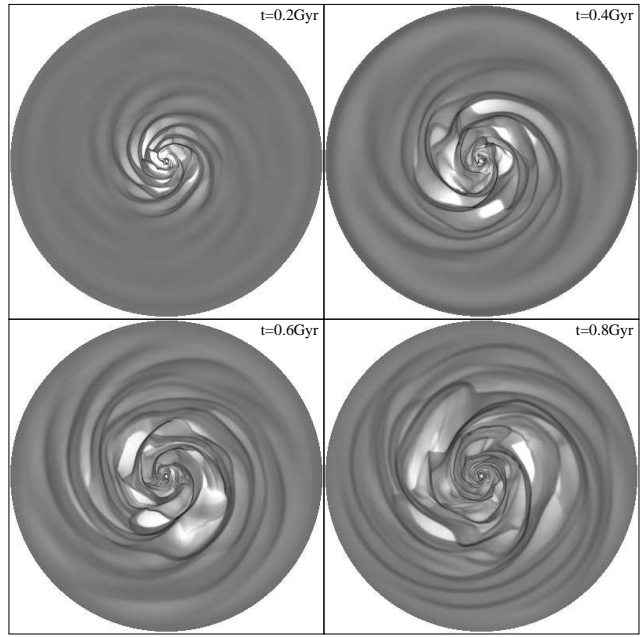,width=84mm}}
\caption{Surface density maps of PPM code calculations of the gas response,
with $a=5$ km s$^{-1}$, to the potential
taken from N-body calculations.  Shade of grey is proportional to $\log(\sigma)$
, with
the darkest regions corresponding to 50 times the initial density, and white 
regions
to $0.01$ times the initial density.  The interval 
between frames is approximately
one rotation at $R=10$ kpc and the disc radius is $30$ kpc.}
\label{fig:sellwood5_rhopics}
\end{figure}

An animation of the calculation result shows that the spiral shocks tend to 
continually
break and re-join, with the portions at smaller radii rotating more rapidly than
 those
at larger radii.  Where the shocks collide, very high densities can be reached
temporarily.  This continual disruption of the spiral structure originates in the
spiral form of the gravitational potential.  In Figure 3, 
%\ref{sellwood_potpics},
images are shown
indicating the potential value in blue, with the gas density superposed in red.
Only the non-axisymmetric potential is shown, i.e.\ the colour at any point is
proportional to the deviation of the potential from the mean value at that 
radius.
These images show that the major spiral shocks do form in the deepest potential
valleys,
which means that (in this `galaxy') the morphology of the shock structure is
effectively determined by the potential structure\footnote{
We note that this behaviour is in contrast to that expected in the
case of a long lived spiral mode, since the steady state response of the
disc in this case involves an angular offset between the shock and the
potential minimum (Gittins and Clarke 2004)}.
Once again, the effect of a reduced
sound speed is only visible in the outer regions of the galaxy.

\begin{figure}
\centerline{\epsfig{figure=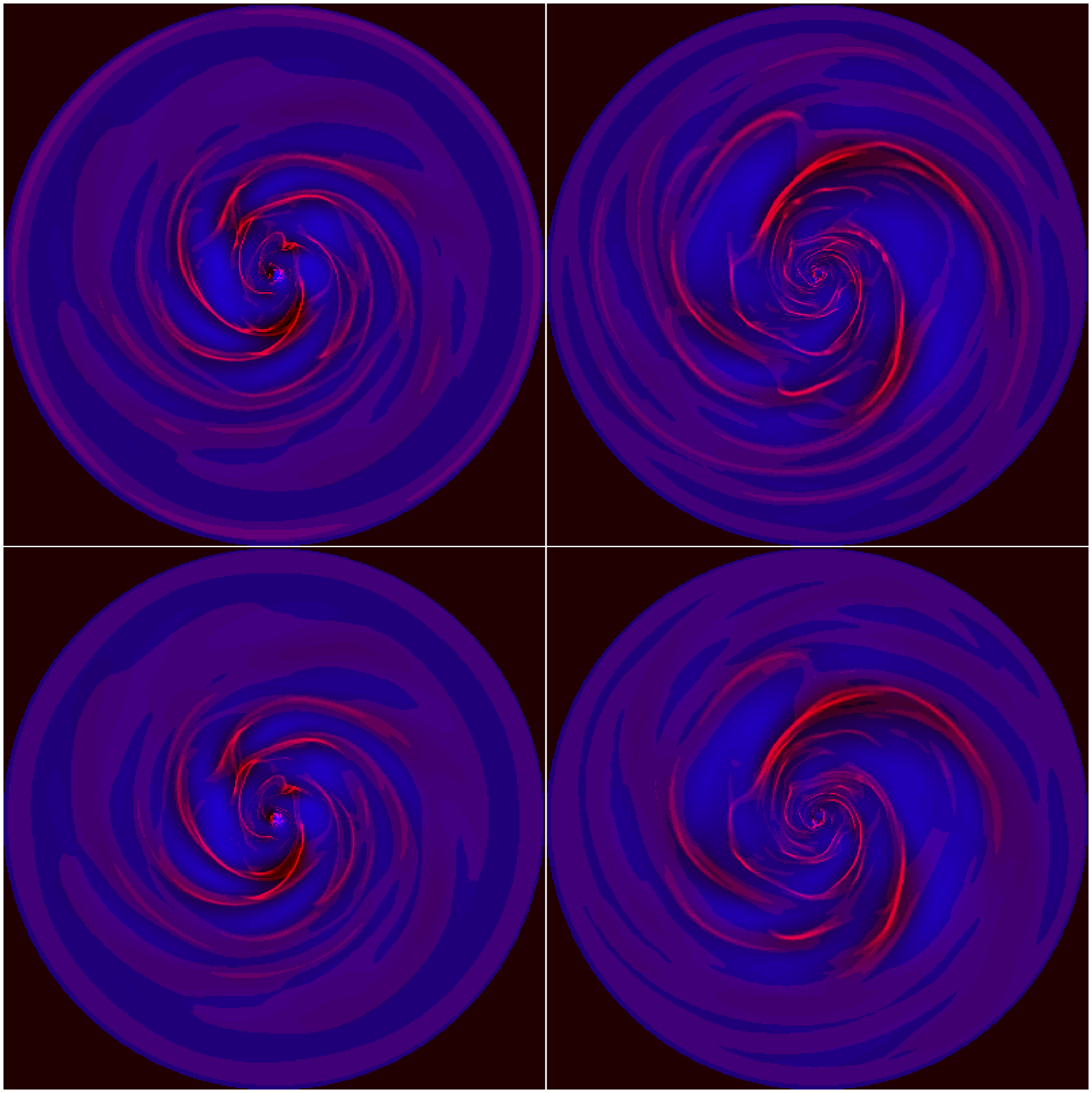,width=84mm}}
\caption{Images of the calculations shown in Figure 
2
%\ref{sellwood5_rhopics},
showing gravitational potential in blue (dark areas correspond to potential 
valleys)
 and gas
surface density in red.  Images are shown at times $t=0.4$ Gyr (left) and $0.8$
Gyr (right),
for the cases $a=5$ km s$^{-1}$ (top) and $a=8$ km s$^{-1}$ (bottom).}
\label{fig:sellwood_potpics}
\end{figure}

The spiral structures formed in the gas disc do not correspond
to one dominating spiral mode, but they do represent clear spiral arms. 
 We have analysed the non-axisymmetric structure in the stellar potential
 by computing the function $C_m$:
 
 \begin{equation}
 C_m = {|\int_0^{2\pi}  \Phi(r,\phi)  e^{-im\phi}\rmn{d} \phi| \over \int_{0}^{2\pi}  \Phi(r,\phi) \rmn{d}\phi}
 \end{equation}
 
 \noindent (where $\Phi$ is the gravitational potential at the mid-plane of the
 N-body simulations) as a function of radius for different snapshots of the
 stellar distribution. $C_m$ represents the {\it local} amplitude of
 the $m$th Fourier mode. The fact that $C_m$ is a function of radius
 at a given time indicates that the structure is not describable
 as a superposition of global modes. In fact, the region of the disc that
 is subject to strong non-axisymmetric disturbances  progressively
 expands outwards as the simulation progresses (Figure 3), 
 as can also be deduced from Figure 2. We find that the maximum value
 of $C_m$ is generally for $m=2$ and $m=3$, with higher order
 modes contributing up to $m=8$ only\footnote{The cut-off in the mode spectrum
 at $m=8$ can readily be understood as corresponding to the minimum
 unstable wavelength for a disc with $Q=1$, for which
 $m\sub{max} = 4 \rmn{tan} i (M\sub{halo}+ M\sub{disc})/M\sub{disc}$ and we estimated
 the pitch angle ($i \sim 30$) from Figure 3.}.
As one would expect, given the variable appearance of the pattern
 in Figures 2 and 3, the relative contribution of the various modes fluctuates
 on a roughly dynamical timescale: for example, the $m=2$ mode is predominant
 at $t=0.4$Gyr (upper
 right panel of Figure 2), whilst at $0.6$ Gyr,
 the $m=3$ mode is of comparable strength and becomes  the dominant mode
 at $t=0.8$ Gyr.

 If such a galaxy were observed, 
the densest regions of gas would correspond to observable features 
 (e.g. enhanced H$\alpha$ emission, concentration of \HII regions or
enhanced surface brightness). There are numerous examples of galaxies
that do not show grand design structure but in which such complex and
multi-armed structure is nevertheless apparent (see, for example, the images
contained in  Elmegreen et al 1992, Fuchs and Moellenhoff 1999). 
Such snapshots give no hint of
the fact that in the simulation, these features are rather short lived
(of order an orbital time).
Structures in the simulations
include arms, spurs, loops and dense `knots', some of which
rotate in the opposite sense to the galaxy.  The interpretation of these
features, observationally, might involve more complex mechanisms such as
self-gravity in the gas, feedback from
star formation or the influence of a galactic companion, but in this case they 
form 
simply from the response of non self-gravitating gas to the
fluctuating stellar potential.

In figure 4,
%\ref{sellwood_cutpics}, 
some images of the densest parts of the galaxy
can be seen.  These are a crude approximation of the shapes of spiral features 
that might
be observed if the calculation were a real galaxy.

\begin{figure}
\centerline{\epsfig{figure=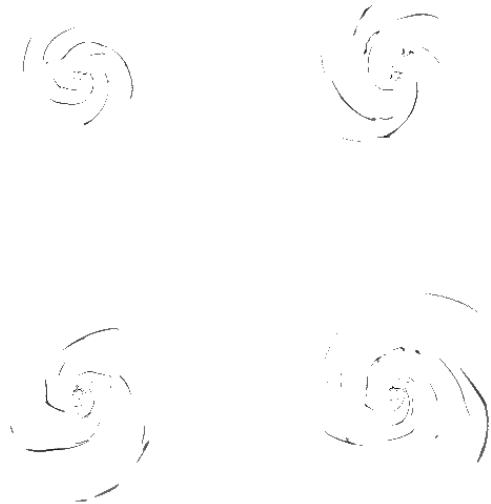,width=84mm}}
\caption{Images of the calculation at various times, showing only gas at densities great
er than
five times the mean.}
\label{fig:sellwood_cutpics}
\end{figure}

\section{Star formation rate variations}
\label{sec:discussion}

The strong time dependence of these features means that the total mass of gas in
 the disc
at high densities also varies significantly with time.
The galaxy could therefore show significant variation
in star formation rate.  In this section, we discuss the
application of the simulations to the star formation history
of spiral galaxies.

\subsection{Star formation history of the simulated galaxy}
\label{sec:SFHsim}
If vigorous star formation is only triggered above a critical
density (as suggested by  Kennicutt 1989) then it is reasonable to quantify the SFR in
terms
of the total mass instantaneously above that density (here denoted
$\Sigma\sub{crit}$).
Following Martin and Kennicutt 2001, we adopt a star formation rate that scales with
the mass above the critical threshold density raised to the power
$1.5$. The time-dependence of the resulting `star formation
rate' is then sensitive to the ratio of the mean gas density to
the threshold density adopted. In general terms, if the mean density 
is close to the threshold, then the variation of star formation rate
is rather smooth. There is an initial period of adjustment (over
$\sim 0.4$ Gyr), when the gas density profile is reconfigured
from the uniform initial conditions to one containing pronounced
spiral structure: the azimuthally averaged density profile remains
roughly uniform during the simulation, 
although the net effect of the torques from the
non-axisymmetric stellar distribution results in a modest (factor
three) enhancement in the azimuthally averaged density in the innermost
regions (within $1$ kpc). Thereafter, in the case that the threshold
density is close to the initial density, the variations in global
star formation rate are rather modest (less than $50 \%$).

  If, however, the critical density exceeds the mean density by a
larger  factor, the resulting star formation variations become
much more pronounced, as bursts of star formation are associated
only with epochs when the form of the spiral pattern in the gas results in
regions of  particularly high density. (For example, the collision
of two spiral features can result in such localised regions of
extreme enhancement). If the critical density is set to five
times the mean density, then the star formation rate exhibits 
peaks   
on roughly the orbital timescale of around 200 Myr: in this simulation
the global star formation rate can increase by a factor of about
$3$ over $\sim 50$ Myr. More extreme variations are seen as the
critical threshold is increased further: Figure 5 shows the 
mass of gas above $\Sigma\sub{crit}$ in the case that the critical
density threshold is $10$ times the initial density 
in the calculation with $a=5$ km s$^{-1}$.
%Figure \ref{sellwoodSFrate} illustrates this idea, showing
%the total mass above a threshold density of 10 times the initial density vs.\ ti
%me
%Significant variation on roughly the orbital timescale of around 200 Myr is seen
Figure 6 
%\ref{sellwoodSFrate_pics} 
shows the surface density of the disc, with the
area above the density threshold highlighted in red, at
a minimum and a maximum of the high-density mass (indicated with 
arrows on Figure 5).
%\ref{sellwoodSFrate}).

\begin{figure}
\centerline{\epsfig{figure=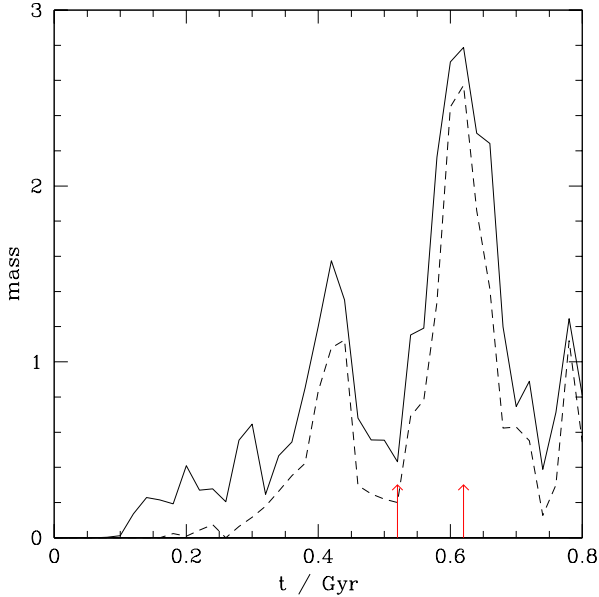,width=84mm}}
\caption{Total mass (arbitrary units, in which the total mass of the gaseous disc is 
$\sim 3000$)
of gas above a threshold density of 10 times the initial density vs.\ time in 
the calculation
with $a=5$ km s$^{-1}$ at radii $R>2$ kpc (solid line) and $R>5$ kpc (dashed line).
Red arrows indicate the two time instants displayed in Figure~6}
%\ref{sellwoodSFrat
%e_pics}.}
\label{sec: sellwoodSFrate}
\end{figure}

\begin{figure}
\centerline{\epsfig{figure=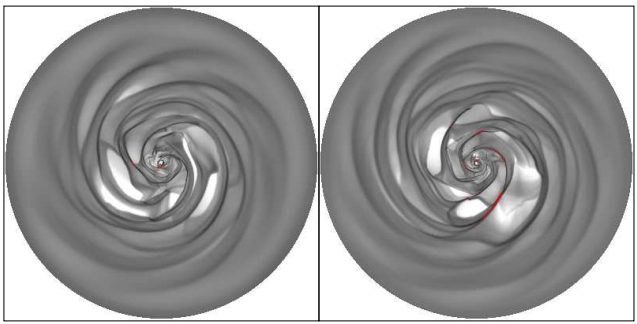,width=84mm}}
\caption{The gas density at $t=0.51$ Gyr (left) and $t=0.61$ Gyr (right), corresponding
to low and high masses at surface density $\sigma > 10 \bar{\sigma}$ 
(highlighted
in red)
as indicated by red arrows on Figure 5} 
%\ref{sellwoodSFrate}.}
\label{sec:sellwoodSFrate_pics}
\end{figure}

The simulations thus indicate that an
isolated galaxy can develop a time dependent spiral structure in the
gas which
in turn may lead to a significantly varying star formation rate.
The amplitude of this variation is strongly dependent on the 
ratio of the mean density in the galaxy to the critical
threshold density, with large amplitude variations in
{\it global} SFR being  
associated with the case that the average surface density is
significantly less than the threshold value. 
The predominant timescale is roughly the orbital timescale at the
galaxy's half mass radius, although there are evidently smaller scale
fluctuations on shorter timescales. The variations are aperiodic, with the
burst amplitude varying strongly from burst to burst.

  We note that  fluctuations in star formation rate have also been
recorded in the N-body/hydrodynamical simulations of Michel-Dansac
and Wozniak 2004,
although comparison with the present case is weakened by the fact that
these simulations do not include a dark halo component, and hence
develop a strong bar. In  the  Michel-Dansac
and Wozniak 2004 simulations, star formation
rate variations are mainly confined to the inner regions of the disc
and result from time-dependent inflow along the bar, in contrast
to the present case where regions of intense star formation are
distributed in the disc, and correspond to events such as the merger
of spiral arms. Another recent simulation that traced the evolution of
the star formation rate is that of Semelin and Combes 2002 . 
However, in this case
the spiral structure does not persist beyond about a galactic rotation
period, this smoothing out being attributed by the authors to the effect
of numerical heating by discrete dark matter particles. This being the
case, it is unsurprising that the  Semelin and Combes simulations
show only smooth variations
of the star formation rate after the first few hundred Myr.
The simulations of Bottema 2003 are probably those which bear the closest
comparison with the ones presented here, since they model non-barred
galaxies in which the stellar potential exhibits transient but regenerative
spiral structure as in the input (Sellwood and Carlberg 1984) potential
employed here. In the Bottema 2003 simulations, however, the resulting
star formation history depends critically on the degree of supernova
feedback employed. In the absence of feedback, dense regions undergo
runaway collapse (an outcome not possible, of course, in our
non self-gravitating simulations), whereas the inclusion of feedback
(modeled as localised injection of thermal energy) not only reduces
the overall star formation level but also diminishes the
amplitude of variations in the
star formation rate, as star formation is self-quenched in dense
regions.

There is evidently considerable uncertainty in how to
model the relation between structures generated in the gas and the
corresponding star formation history, when the critical
processes (star formation and feedback) occur on a scale far smaller
than the resolution elements of current simulations. 
For example,
 Li et al (2005) have modeled star formation in unstable disc
 galaxies through the introduction of hydrodynamic `sink particles',
 at a mass scale of around $\sim 10^7 M_\odot$, and have quantified
 the star formation rate in terms of the mass accumulated in such
 sink particles. These simulations also show a rather bursty star
 formation history; as in the simulations we report here, the stellar
 component of the galactic disc is not cooled efficiently and the spiral
 structure in the stars is very weak after a Gyr. Since this washing
 out of spiral structure proceeds from the inside out, and since the
 gas surface density distribution is exponential (rather than uniform
 as in the simulations here) this means that the burstiness of
 the star formation decreases more strongly than in our
 simulations, where the response is more heavily weighted
 towards conditions in the outer disc. We emphasise, however, that
 if the stellar population is subject to effective cooling, regenerative
 spiral structure persists over many Gyr at all disc radii (Sellwood
 and Carlberg 1984) and hence one would expect the bursty star
 formation history to be likewise a persistent phenomenon.
 
  In summary, we emphasise the possible 
 role of regenerative spiral structure in producing
global variations in star
formation rate,
we consider that the expected {\it amplitude} of such star formation
rate  variations
is far from clear.

\subsection{Observational Constraints on SFR variations in disc galaxies}
\label{sec:SFHobs}

  Given a stellar population where each star can be placed in an HR
diagram, it is in principle possible to place some constraints on
the star formation history (SFH) that is compatible with the observed
distribution. Several authors have attempted this in the solar
neighbourhood,  e.g. Rocha-Pinto et al 2000, Hernandez et al 2000,
Gizis et al 2002;
the former two papers found evidence for approximately
periodic star formation `bursts' (involving enhancements in the
SFR by a factor $2-4$) with typical
spacing  $\sim 0.5$ Gyr, i.e. comparable with the local galactic rotation
period. The latter authors did not however find any evidence for such bursts
over the last $4$ Gyr.

  In external galaxies, SFR variations may be indirectly constrained through
the application of population synthesis models to integrated stellar
spectra, 
 e.g. Kauffmann et al  2003. The absorption line indices used in this
study are sensitive to isolated bursts during the last $1-2$ Gyr, but
it is unclear whether they  would be sensitive to the sort of continuously
fluctuating star formation rate over this period as depicted in
Figure 5. Likewise, element ratios may be used as a diagnostic of
a bursty star formation history (Gilmore and Wyse 1991) but would again be
insensitive to the SFH in Figure 5, since enhancement of iron to
$\alpha$ element ratios requires an extended (many Gyr) period of
suppressed star formation. On the other hand, in the case of galaxies whose
mean gas surface density is more than an order of magnitude below the
threshold value, the SFH takes the form of discrete and well separated
bursts which could leave an imprint on element ratios.

 Another technique, in relatively nearby galaxies,
is to trace {\it local} variations in SFR over the last
few hundred  Myr through photometric maps of resolved stellar populations.
Unlike solar
neighbourhood studies, or population synthesis models of entire galaxies,
such mapping has the potential to disentangle  local and global effects
and could, in principle, distinguish between density wave theories
(where a given location is periodically over-run by a star formation
triggering event) or models involving material arms that corotate
with the galaxy locally. Williams 2003 found evidence for temporal
SFR variations at a given location but nevertheless found that there
are regions with consistently higher SFRs than others over the time
sampled. In our models, the interplay of spatial and temporal effects
is complex: we expect neither the result of spiral density wave theory
(where each region undergoes a phase lagged SFR variation of equal
amplitude) nor that of rigid material arms, in which case each region is
permanently in a state of high or low SFR. In our models, by contrast,
the arms cross the disc but are themselves evolving on a similar
timescale.  Figure 7
%\ref{histpics} 
shows the surface density vs.\ time for
two different small patches in the disc, traced along circular orbits over
the entire length of the calculation.  We see that although some
regions  may pass through several arms with no significant
activity (right hand panel), in  other regions there is a continual variation
in density. In the latter case (left hand panel), we would expect significant
star formation variations even if the threshold density  were set equal to the
mean surface density (unity in these calculations). Thus we conclude  that
although large amplitude variations in {\it global} SFR require the
mean density to be less than $\Sigma\sub{crit}$, the SFR will show strong {\it local} 
variations even when the mean density is similar to $\Sigma\sub{crit}$.

\begin{figure*}
\centerline{\epsfig{figure=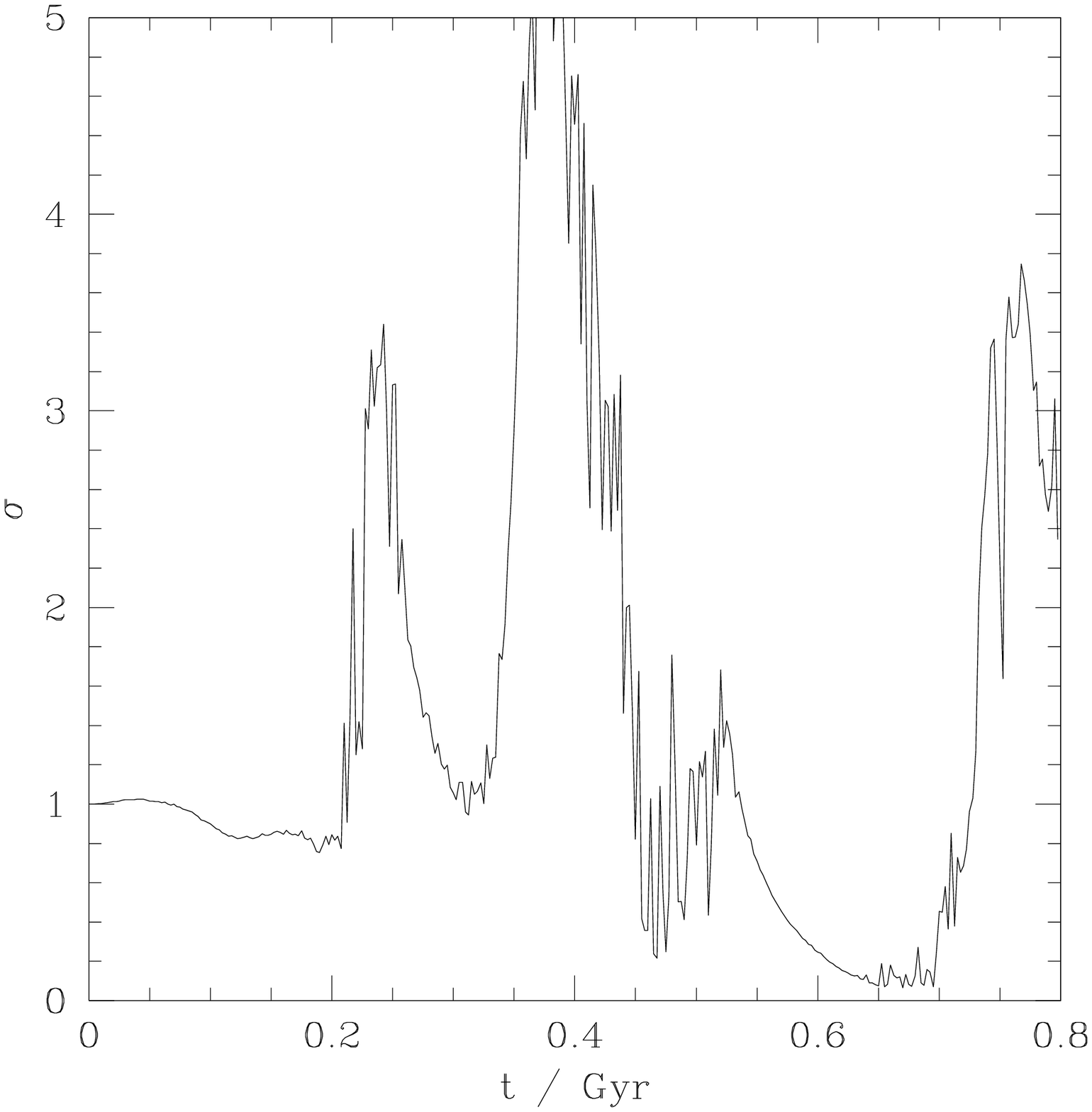,width=84mm}
            \epsfig{figure=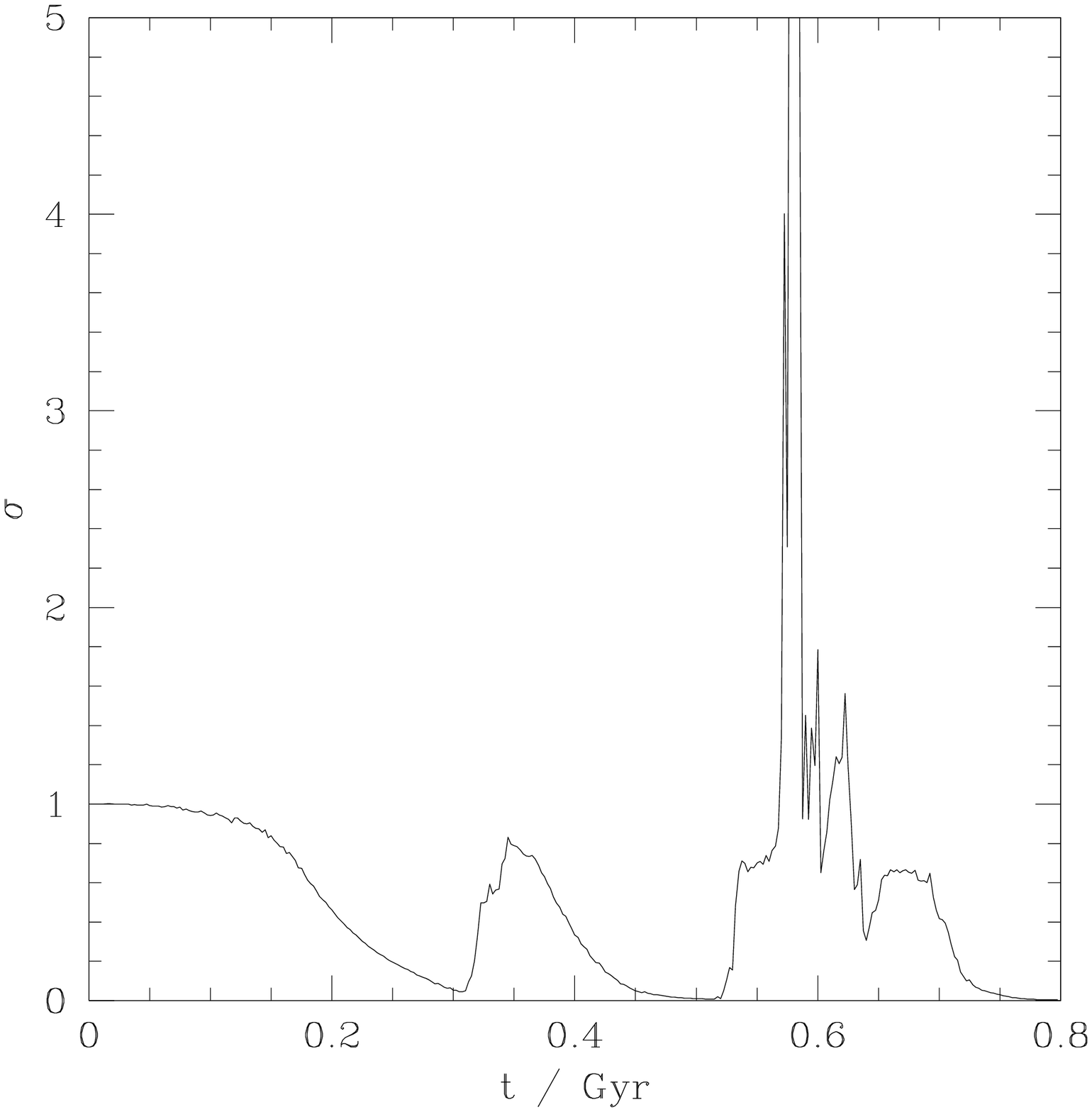,width=84mm}}
\caption{ Surface density vs.\ time for two small regions of gas 
at 10 kpc traced along
 their orbits.  
 The first panel shows a region undergoing a high level of 
 activity throughout,
whereas the second shows a region that remains relatively quiescent until later
in the
calculation.}
\label{sec: histpics}
\end{figure*}

\section{Conclusions}
\label{sec:conclusions}

We have presented the response of an isothermal gas disc to the gravitational
potential of an N-body simulation of an isolated spiral galaxy.  Complex spiral
structures are formed, which break and re-join continually on a timescale comparable
with the local
orbital period.  This highly varying
structure leads to the formation of many dense features, which are generally 
short-lived.
This would be expected to correspond to  variations in the global star formation
rate, as events such as the merging of spiral features occur sporadically
on timescales of a few  galactic rotation periods. Since the amplitude
of such bursts varies from burst to burst, it is possible that rarer, larger
scale bursts (traditionally attributed to galactic interactions) may also
occur in {\it isolated} disc galaxies, due purely to the regenerative
nature of spiral structures in stellar discs.
We however stress that large scale variations in {\it global} SFR are probably
confined to galaxies with mean gas surface 
density considerably below the threshold value.

 The simulations also imply spatial and temporal
variations in star formation rates that are considerably more complex
than those generated by a long lived spiral mode, since now
the lifetime of individual spiral features is comparable with
the timescale on which they cross the disc. Mapping of resolved
stellar populations in nearby galaxies offers a potential
tool to examine the observational evidence for such effects
(Kodaira et al 1999, Williams 2003).

  We stress that our simulations omit many processes that {\it must}
be important in real galaxies (e.g. interconversion between the star
and gas phase,  the effects of stellar feedback,
self-gravity and equation of state of the gas and back reaction of the gas
on the stellar dynamics). For this
reason, the results in this paper are more illustrative than quantitative
(and indeed, as we discuss in 4.1, 
%\ref{SFsim}, 
it is not obvious how to
quantify star formation variations in {\it any}  simulation where the scales
on which star formation and feedback occur are so far below the resolution
of the simulations).

 Finally, we note that N-body simulations (such as those of Sellwood and
Carlberg 1984),
which find spiral structures to be recurrent, short-lived patterns,
have attracted much interest from those studying the origin of spiral
arms in galaxies. It is only rather recently that some of the
{\it observational}  signatures of such transient structures have
begun to be explored (Sellwood and Preto 2002,
Sellwood and
 Binney 2002).  Irregular variations in star
formation rate are an obvious corollary of such a picture. The study
of temporal and spatial variations in star formation rates
in nearby galaxies may thus shed some light on
the nature and origin of spiral structure in disc galaxies.

\section*{Acknowledgments}

Thanks are due to Jerry Sellwood for performing the N-body simulation of the 
galaxy and
supplying the potential data, to Peter Teuben for providing the CMHOG code and
 support,
and to Rob Kennicutt, Max Pettini and Gerry Gilmore 
 for advice on star formation histories.
 We also thank Giuseppe Lodato for useful input on the mode spectrum
 in gravitationally unstable discs. CJC acknowledges 
 support from the Leverhulme Trust in the form of a Philip Leverhulme Prize.
We are grateful to the referee (Curtis Struck) for encouraging us to
examine how our results depend on the ratio of threshold to mean
gas surface density and for other comments that have improved the paper.
\bibliographystyle{mn2e} 
\bibliography{refs}

\end{document}